\documentclass[]{aa}
\usepackage{graphicx}
\usepackage{amsmath}
\usepackage{natbib}

\begin{document}

\headnote{Research Note}
\title{Signatures of highly inclined accretion disks in Galactic Black Hole Candidates and AGNs}
\author{Zakharov A.F.\inst{1,2,3} \and Repin S.V.\inst{4}}
\offprints{Zakharov A.F., \email{zakharov@vitep1.itep.ru}}
\institute{National Astronomical Observatories, Chinese Academy of Sciences, 100012,
Beijing, China,
\and Institute of Theoretical and Experimental Physics,
           25, B.Cheremushkinskaya st., Moscow, 117259, Russia,
\and Astro Space Centre of Lebedev Physics Institute, 84/32,
Profsoyuznaya st.,
             Moscow, 117810, Russia,
\and Space Research Institute, 84/32,
Profsoyuznaya st.,
             Moscow, 117810, Russia,\\
}
\date{Received / accepted }

\abstract{
     Recent X-ray observations of microquasars and Seyfert galaxies
reveal the broad emission lines in their spectra, which can arise
in the innermost parts of accretion disks. Simulation indicates
that at low inclination angle the line is registered by distant
observer with characteristic two-peak profile. However, at high
inclination angles (  $> 85 \degr$) two additional peaks arise.
This phenomenon was discovered by \citet{MPS93} using the
Schwarzschild black hole metric to analyze  such effect. They
assumed that the effect is applicable to a Kerr metric far beyond
of a range of parameters that they exploited. We check and
confirm their hypothesis about such structure of the spectral
line shape for a Kerr metric case. We use no astrophysical
assumptions about physical structure of the emission region
except assumption that the region should be narrow enough.
 Positions and heights of these extra peaks drastically
depends on both the radial coordinate of the emitting region
(circular hot spot) and the inclination angle. We find that best
conditions to observe this effect are realized at $\theta >
85\degr$ and $r > 5r_g$ and may exist in microquasars or low-mass
black holes in X-ray binary systems, because there is some
precession (and nutation of accretion disks) with not very long
time periods (see, for example, SS433 binary system). The line
profiles for different inclination angles and radial coordinates
are presented. To analyze  an influence of disk models on the
spectral line shapes we simulate the line profiles for Shakura --
Sunyaev disk model for accretion disks with the high inclination.

 }
 \authorrunning{Zakharov \& Repin}
 \titlerunning{Signatures of highly inclined accretion disks }

\maketitle
\section{Introduction}

More than ten years ago it was predicted that profiles of lines
emitted by AGNs and X-ray binary systems\footnote{Some of them are
microquasars (for details see, for example, \cite{Grein99, Mira00,
Mira02a}).} could have asymmetric double-peaked shape (e.g.
\cite{Chen89,Fabian89,Robinson90,Dumont90,MPS93}). Generation of
the  broad $K_\alpha$ fluorescence lines as a result of
irradiation of cold accretion disk was discussed by many authors
(see, for example,
\cite{MPP91,MPPS92,MPPS92a,Matt92,MFR93,Bao93,Mart02} and
references therein).
    Recent X-ray observations of Seyfert galaxies, microquasars
and binary systems
(\cite{fabian1,tanaka1,nandra1,nandra2,malizia,sambruna,
yaqoob4,ogle1,Miller02} and references therein) confirm these
considerations in general and reveal broad emission lines in their
spectra with characteristic two-peak profiles. A comprehensive
review by \cite{Fabi00} summarizes the detailed discussion of
theoretical aspects of possible scenarios for generation of broad
iron lines in AGNs. These lines are assumed to arise in the
innermost parts of the accretion disk, where the effects of
General Relativity (GR) must be taken into account, otherwise it
appears very difficult, if any, to find the natural explanation
of the observed line profile.

     Numerical simulations of the line structure
could be found in a number of papers:
(\cite{Koji91,Laor91,Bao92,Bao93,BHO94,rauch,Rauch94,bromley,Fan97,pariev2,pariev1,pariev3,
Rusz00,Ma02}). They indicate that the accretion disks in Seyfert
galaxies are usually observed at the inclination angle $\theta$
close to $30\degr$ or less. It occurs because according to the
Seyfert galaxy models, the opaque dusty torque, surrounding the
accretion disk, so, such structure does not allow us to observe
the disk at larger inclination angles.

     However, at large inclination angles $\theta > 80\degr$ the new
observational manifestations of GR could arise. (\cite{MPS93}
discovered such phenomenon for a Schwarzschild black hole,
moreover the authors predicted that their results could be
applicable  for a Kerr black hole over the range of parameters
which were exploited). The authors mentioned that this problem
was not analyzed in details for a Kerr metric case and it would be
necessary to investigate this case. Below we do not use some
specific model on surface emissivity of accretion (we only assume
that the emitting region is narrow enough). But general statements
(which will be described below) could be generalized on a wide
disk case without any problem. Therefore, in this paper we check
and confirm their hypothesis for Kerr metric case and for a
Schwarzschild black hole using another assumptions about surface
emissivity of accretion disk. In principle, such phenomenon could
be observed in microquasars and X-ray binary systems where there
are neutron stars and black holes with stellar masses.

     In Subsection $3\degr$ we discuss  disk models used for
simulations. In Subsection $4\degr$ we present the results of
simulations with two extra peaks in the line profile. These extra
peaks exist because of the gravitational lens effect in the
strong field approximation of GR. In Subsection $5\degr$ we
discuss results of calculations and present some conclusions.

\section{Numerical methods}

We used an approach which was discussed in details in papers by
\cite{zakh91,zakharov1,zak_rep1,zak_rep2,zak_rep02a,zak_rep03,ZKLR02}.
The approach was used in particular to simulate   spectral line shapes.
For example, \cite{ZKLR02} used this approach to simulate an
influence of magnetic field on spectral line profiles. This
approach is based on results of qualitative analysis (which was
done by \cite{zakh86,zakh89} for different types of geodesics near
a Kerr black hole).
 The equations of photon motion in Kerr metric are reduced to
the following system of ordinary differential equations
in dimensionless Boyer -- Lindquist coordinates
(\cite{zakh91,zakharov1,zak_rep1}):
\begin{eqnarray}
   \frac{dt}{d\sigma}
                           & = &
      - a \left(a \sin^2\theta - \xi\right) +
      \frac{r^2 + a^2}{\Delta}
       \left(r^2 + a^2 - \xi a\right),
                        \label{eq1}                       \\
   \frac{dr}{d\sigma} & = & r_1,       \label{eq2}        \\
   \frac{dr_1}{d\sigma}    & = &
      2r^3 + \left(a^2 - \xi^2 - \eta\right) r +
      \left(a - \xi\right)^2 + \eta,                      \\
   \frac{d\theta}{d\sigma} & = & \theta_1,                \\
   \frac{d\theta_1}{d\sigma}
                           & = &
      \cos\theta \left(\frac{\xi^2}{\sin^3\theta} -
                       a^2 \sin\theta
                 \right),              \label{eq5}       \\
   \frac{d\phi}{d\sigma}   & = &
      - \left(a - \frac{\xi}{\sin^2\theta}\right) +
      \frac{a}{\Delta}
           \left(r^2 + a^2 - \xi a \right),
                     \label{eq6}
\end{eqnarray}
 where $\Delta=r^2-2r+a^2$; $\eta = Q/M^2E^2$ and $\xi = L_z/ME$
are the Chan\-dra\-sekhar constants (\cite{chandra}) which are
derived from the initial conditions of the emitted quantum in the
disk plane. The system~(\ref{eq1})-(\ref{eq6}) has two first
integrals
\begin{eqnarray}
  \epsilon_1 & \equiv & r_1^2 - r^4 -
      \left(a^2 - \xi^2 - \eta\right) r^2 -    \nonumber \\
            &        &
      \phantom{.} -
      2\left[\left(a - \xi\right)^2 + \eta \right] r +
      a^2\eta = 0,                  \\
  \epsilon_2 & \equiv & \theta_1^2 - \eta - \cos^2\theta
      \left(a^2 - \frac{\xi^2}{\sin^2\theta}\right) = 0,
                    \label{eq8}
\end{eqnarray}
which can be used for the accuracy control of computation.

     Solving Eqs. (\ref{eq1})--(\ref{eq6}) for
monochromatic quanta emitted by a ring we can calculate a specral
line shape $I_\nu(r,\theta)$ which is registered by a distant
observer at inclination angle $\theta$.

     The numerical integration has been performed using
the combination of the Gear and Adams methods (\cite{gear}) and
realized as the standard package by
\cite{hindmarsh,petzold,hiebert}. We obtain the entire disk
spectrum by summation of sharp ring spectra.

\section{Disk  model}

    To simulate the structure of the emitted line it is necessary
first to choose a model for an emissivity of an accretion disk. We
exploit two different models, namely  we consider narrow and thin
disk moving in the equatorial plane near a Kerr black hole as the
first model and as  we analyze the inner wide part of accretion
disk with a temperature distribution which is chosen according to
the \cite{shasun} with fixed inner and outer radii $r_i$ and $r_o$
as the second model.  Usually a power  law is used  for a  wide
disk emissivity (see, for example,
\cite{Laor91,MPP91,MM96,MKM00,MMK02}). However, another models for
emissivity could not be forbidden for so wide class of accreting
black holes, therefore, just to demonstrate how another emissivity
law could change line profiles we exploit  such emissivity law.

So, at the first, we assume that the source of the emitting quanta
is a narrow and thin disk rotating in the equatorial plane of a
Kerr black hole. We also assume that the disk is opaque for
radiation, so that a distant observer situated on one disk side
cannot register the quanta emitted from its other side.

    For the sake of computational simplicity we suggest that the
spectral line is monochromatic in its co-moving frame. It can
really be adopted because even at $T = 10^8$~K the thermal line
width
\begin{equation}
    \frac{\delta f}{f} \sim \frac{v}{c} \sim
     \frac{1}{c}\sqrt{\frac{kT}{m_{_{\rm Fe}}}} \approx 10^{-3}
     \nonumber
\end{equation}
appears to be much less than the Doppler line width associated
with the Kepler velocity of the disk rotation.

For second case of the Shakura -- Sunyaev disk model, we assume
that a local emissivity is proportional to surface element and
and $T^4$, where $T$ is a local temperature.
     The emission intensity of the ring is proportional to its area.
The area of emitting ring (width $dr$) differs in the Kerr
metric from its classical expression $dS = 2\pi rdr$  and should
be replaced with
\begin{equation}
   dS = \frac{2\pi \left(r^2 + a^2\right)}
             {\displaystyle\sqrt{r^2 - rr_g + a^2\phantom{1}}}\, dr.
          \label{eq9}
\end{equation}

    Thus, the total flux density emitted by the disk and
registered by distant observer is proportional to the integral
\begin{equation}
   J_\nu(\theta) = \int\limits_{r_{in}}^{r_{out}}
              I_\nu(r,\theta) T^4(r) dS,
          \label{eq10}
\end{equation}
where $I_\nu(r,\theta)$ is obtained from the solution of
equations (\ref{eq1})--(\ref{eq6}), $T(r)$ -- from the
appropriate dependence for a hot disk and $dS$ -- from
Eq.~(\ref{eq9}).

     For simulation we assume that
the emitting region lies entirely in the innermost region of
$\alpha$-disk (zone~$a$) from $r_{out} = 10\,r_g$ to
$r_{in} = 3\,r_g$ and the emission is monochromatic in the
co-moving frame.\footnote{We use as usual the notation
$r_g=2GM/c^2$.}
 The frequency of this emission set as a unity
by convention.

\section{Simulation results}

\begin{figure*}
\begin{center}
\includegraphics[width=0.8\textwidth]{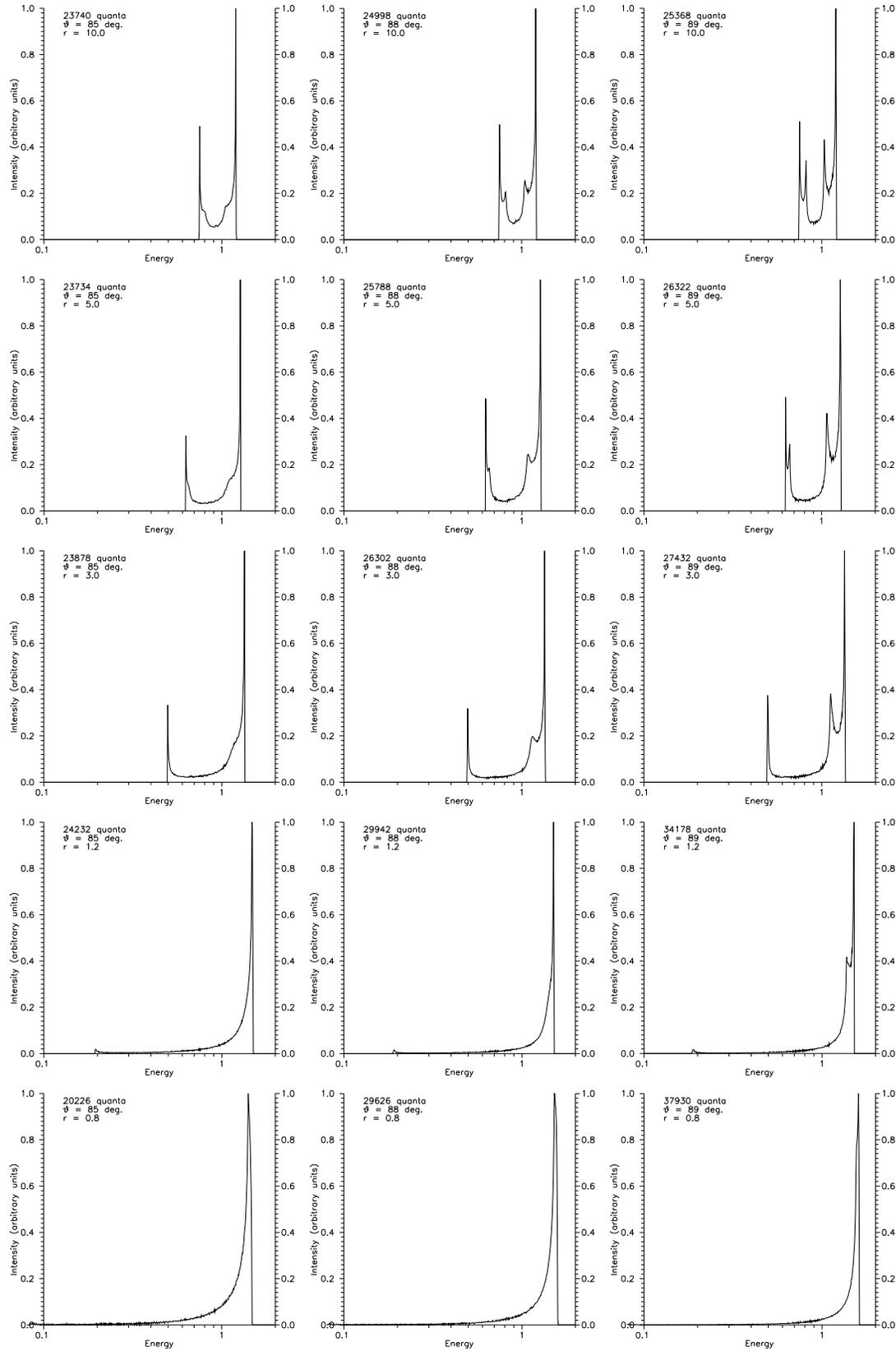}
\end{center}
  \caption{Line profiles with specific details, which arise for
           high inclination angles $\theta > 85\degr$ due
           gravitational lens effect in the strong gravitational
           field approach. The Kerr (rotation) parameter was
           chosen like $a = 0.9981$. The radii decrease
           from top to bottom, the inclination angles increase from
           left to right; their values are shown in each panel along
           with the number of quanta included in spectrum.}
  \label{peaks01}
\end{figure*}

      Spectral line profiles of a narrow ring observed at large inclination
angles $\theta > 85\degr$ and different radii $r$ are shown in
Fig.~\ref{peaks01} (one could say
 that is a Kerr metric generalization of Fig. 1 from \cite{MPS93} which was
drawn using calculations for the Schwarzschild case). The ring is
assumed to move in the equatorial plane of a Kerr black hole with
almost extreme rotation parameter  $a = 0.9981$. The inclination
angle increases there from left to right and radial coordinate --
from bottom to top. The figure indicates that practically there
are no new specific features of profiles, thus,  the line profile
remains one-peaked with a maximum close to $1.6\,E_{lab}$ and
very long red wing without any significant details.
 For lowest radii there are no signatures of multiple peaks of
spectral line shapes even for high inclination angles (the bottom
raw in Fig. 1 which corresponds to $r=0.8\,r_g$).

   With increasing the radius to $r = 1.2\, r_g$ the additional
blue peak arises in the vicinity of the blue maximum at the
highest inclination angle $\theta = 89\degr$. The red maximum
is so small that no details can be distinguished in its
structure. At lower inclination angles $\theta \le 88\degr$
the blue maximum also has no details and the entire line
profile remains essentially one-peaked.

    For $r = 3\, r_g$ the additional detail in the blue peak
appears for $\theta \ge 85\degr$. Thus, for $\theta = 85\degr$ we
have a fairly distinguished bump, for $\theta = 88\degr$ it
changes (transforms) into a small complementary maximum and for
$\theta = 89\degr$ this maximum becomes well-distinguished. Its
position in the last case differs significantly from the main
maximum: $E_3 = 1.12\,E_{lab}$, $E_4 = 1.34\,E_{lab}$.

    With further increasing the radius the red maximum
bifurcates also. This effect becomes apparently visible  for $r =
5\,r_g$ and $\theta \ge 88\degr$.  Thus, for $\theta = 85\degr$ we
have only a faintly discernible (feebly marked) bump, but for
$\theta = 88\degr$ both complementary maxima (red and blue)
arise. For $\theta = 89\degr$ we have already four maxima in the
line profile: $E_1 = 0.63\,E_{lab}$, $E_2 = 0.66\,E_{lab}$, $E_3 =
1.07\,E_{lab}$, $E_4 = 1.28\,E_{lab}$. Note that the splitting
for the blue and red maxima is not equal, moreover, $E_4 - E_3
\approx 7\, \left(E_2 - E_1\right)$.

    For $r = 10\, r_g$ the profile becomes more narrow,
but the complementary peaks appear very distinctive. We have
four-peak structure for  $\theta \ge 88\degr$. It is interesting
to note that for $\theta = 89\degr$ the energy of the blue
complementary peak is close to its laboratory value.

   Note that the effect almost disappears when the radial coordinate
becomes less than $r_g$, i.e. for the orbits which could exist
only near a Kerr black hole.

\begin{figure*}
\begin{center}
\includegraphics[width=0.98\textwidth]{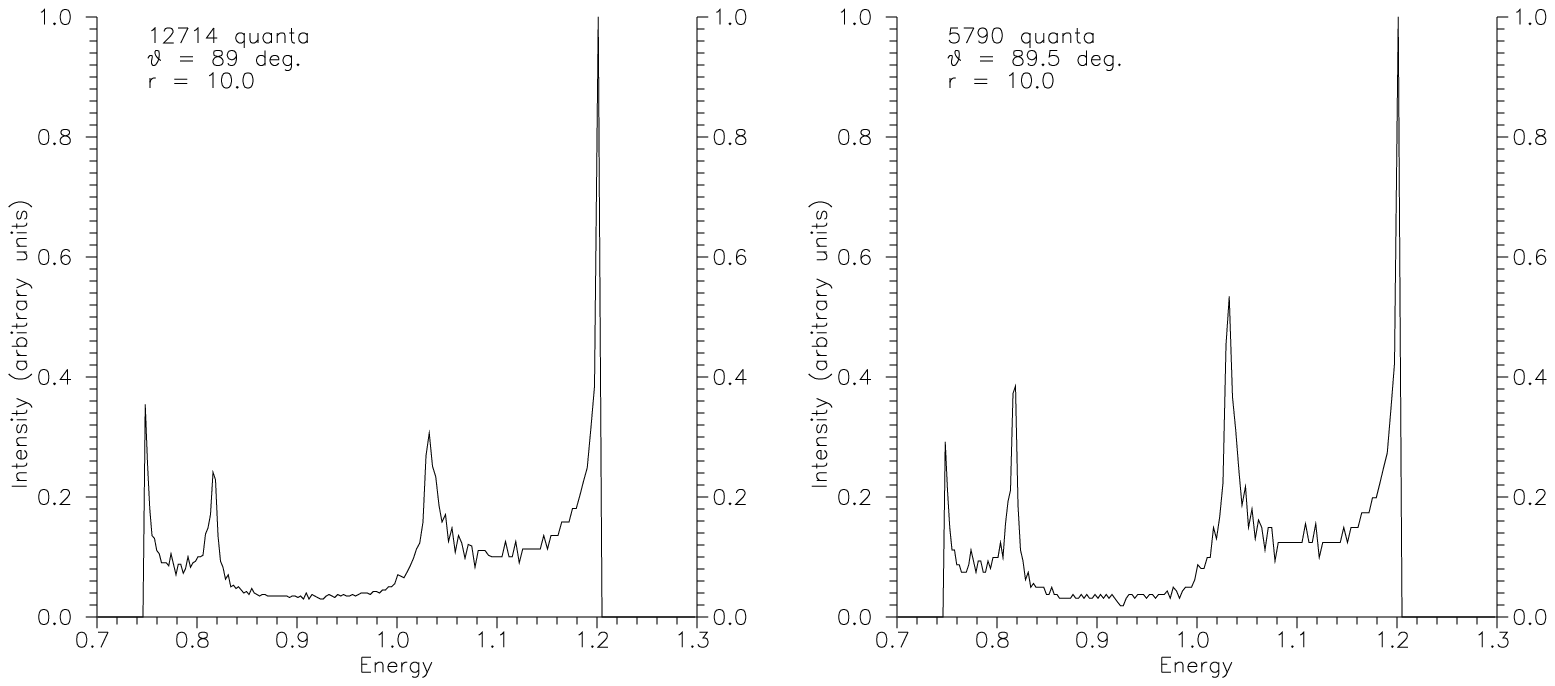}
\end{center}
  \caption{Details of a hot spot line profile for the most
           distinctive case with $r=10\,r_g$ and $a = 0.9981$
           (see the top row of Fig.~\ref{peaks01}).
           The images of all orders are counted. Left panel includes
           all the quanta, registered at infinity with $\theta > 89\degr$.
           Right panel includes the quanta with $\theta > 89\fdg5$.}
  \label{peaks02}
\end{figure*}

\begin{figure*}
\begin{center}
\includegraphics[width=0.98\textwidth]{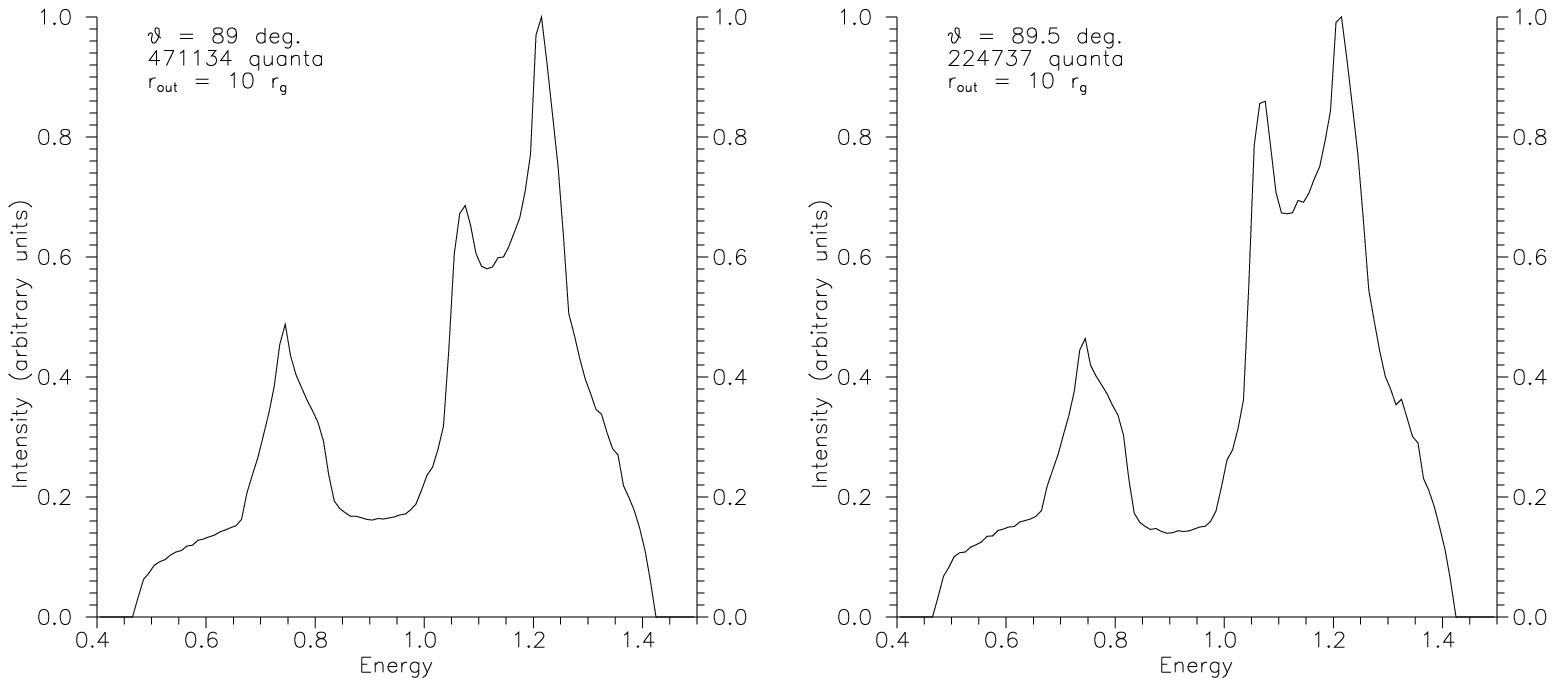}
\end{center}
  \caption{Details of a line structure for $\alpha$-disk in
           the Schwarzschild metric with outer and inner edges of
           emitting region equal to $r_{out} = 10\,r_g$ and
           $r_{in} = 3\,r_g$, respectively. Left panel includes all
           the quanta, registered at infinity with $\theta > 89\degr$.
           Right panel includes the quanta with $\theta > 89\fdg5$.}
  \label{peaks03}
\end{figure*}

   Fig.~\ref{peaks02} demonstrates the details of the spectrum
presented in the top row in Fig.~\ref{peaks01}. Thus, the left
panel includes all the quanta emitted by a hot spot at $r =
10\,r_g$ with $\theta > 89\degr$ at infinity (the mean value could
be counted as $85\fdg5$ if the quanta distribution  would be
uniform there) but with much higher resolution than in
Fig.~\ref{peaks01}. The spectrum has four narrow distinctive
maxima separated by lower emission intervals. In the right panel
which includes all the quanta with
 $\theta > 85\fdg5$ the right complementary maxima is even higher
than the main one. The blue complementary maxima remains still
lower than its main counterpart, but it increases rapidly its
intensity with increasing the inclination angle. It follows
immediately from the comparison of left and right panels. Some
"oscillation behavior" of the line profile between the maxima has
a pure statistical origin and does not refer to physics.

   As an illustration a spectrum of an entire accretion disk at
high inclination angles in the Schwarzschild metric is shown in
Fig.~\ref{peaks03}. (In reality we've calculated geodesics for
the quanta trajectories in the Schwarzschild metric using the
same Eqs. (\ref{eq1})--(\ref{eq6}) as for a  Kerr metric, but
assuming there $a = 0.01$.) The emitting region (from 3 to
$10\,r_g$) lies as a whole in the innermost region of
$\alpha$-disk (the detailed description of this model was given
by \cite{shasun,LipShak}).

   As it follows from the Fig.~\ref{peaks03} the blue peak may consists of the two
components, whereas the red one remains unresolved.

\section{Discussion and conclusions}

    The complicated structure of the line profile at large inclination
angles is explained by the multiple images of some pieces of the
hot ring. One could point out that the result was obtain in the
framework of GR without any extra physical and astrophysical
assumptions about character of radiation etc. For a Kerr black
hole we assume only that radiating ring is circular and narrow.

The problem of multiple images in the accretion disks and extra
peaks has first been considered by (\cite{MPS93}, see also
\cite{Bao92,Bao93,BHO94,BHO96}). Using numerical simulations
\cite{MPS93} proved the statement for the Schwarzschild metric
and suggested that the phenomenon is applicable to a  Kerr metric
over the range of parameters that the authors have analyzed.
\cite{MPS93} noted also that it is necessary to realize detailed
calculations to confirm their hypothesis.  We verified and
confirmed their conjecture without any assumptions about a
specific distribution of surface emissivity or accretion disk
model (see, Fig. \ref{peaks01},\ref{peaks02}).

 We confirmed also a conclusion of \cite{MPS93} that extra peaks
are generated by photons which are emitted by the far side of the
disk, therefore we have a manifestation of gravitational lensing
in the strong gravitational field approach for GR.

Some  possibilities to observe considered features of spectral
line profiles was considered by \cite{MPS93, Bao93}. The authors
argued that there are non-negligible chances to observe such
phenomenon in some AGNs and X-ray binary systems. For example,
\cite{Bao93} suggested that NGC 6814 could be a candidate to
demonstrate such phenomenon  (but \cite{Made93} found that the
peculiar properties of NGC6814 are caused by a cataclysmic
variable like an AM Herculis System).

However, it is clear that in general a probability to observe
objects where one could find such features of spectral line is
small. Moreover, even if the inclination angle is very close to
$90^{\degr}$, a thickness of disk ( shield or a torus around
accretion disk) may not give us a possibility to look at the
inner part of accretion disk.  Let us discuss astrophysical
situations when the the inclination angle of accretion disk is
high enough.

About 1\% of all AGN or microquasar systems could have an
inclination angle of accretion disk  $> 89^{\degr}$. For example,
\cite{Kor96} found that NGC3115 has very high inclination angle
about $81^{\degr}$ (\cite{Kor92} discovered a massive dark object
$M \sim 10^9 M_\odot$(probably, massive black hole) in NGC3115).
Perhaps, we have much higher probability to observe such
phenomenon in X-ray binary systems where black holes with stellar
masses could be. Taking into account an precession which actually
observed for some X-ray binary systems (for example, there is a
significant precession of the accretion disk for SS433 binary
system \cite{Cher02},\footnote{\cite{Shak72} predicted that if
the plane of an accretion disk is tilted relatively to the
orbital plane of a binary system the disk can precess.} moreover
since the inclination of the orbital plane is high (i $\sim
79^{\degr}$ for this object), so, sometimes we could have a
possibility to observe almost edge-on accretion disks for such
objects. Observations indicated that there is  a strong evidence
that optically bright accretion disk in SS433 is in supercritical
regime of accretion. First description for supercritical
accretion disk was given by \cite{shasun}. Even now such model is
discussed in  to explain observational data for SS433
\cite{Cher02}. We used a temperature distribution from Shakura --
Sunyaev model (\cite{shasun,LipShak}) for inner part of accretion
disk to simulate shapes of lines which could be emitted from this
region (Fig. \ref{peaks03}). Therefore, we should conclude that the
properties of spectral line shapes discovered by \cite{MPS93} are
confirmed also for such emissivity (temperature) distributions
which correspond to Shakura -- Sunyaev model.

Thus, such properties of spectral line shapes are robust enough
in respect to wide variations of rotational parameters of black
holes and a surface emissivity  of accretion disks as it was
predicted by \cite{MPS93}. In this work  their conjecture it was
verified and confirmed not only for a Kerr black hole case but
also for another dependence of surface emissivity of accretion
disk.

\begin{acknowledgements}
    One author (A.F.Z.) is grateful to E.F.~Za\-kha\-rova for the
kindness and support, necessary to complete this work. AFZ would
like to thank also Dipartimento di Fisica Universita di Lecce,
INFN, Sezione di Lecce
and National Astronomical Observatories of Chinese Academy of Sciences
 for a hospitality and profs. F.~DePaolis,
G.~Ingrosso, J.Wang and Dr. Ma for very useful discussions.
This work was supported by National Science Foundation of China,
No.:10233050.

    Another author (S.V.R.) is very grateful to
Prof.~E.Sta\-ro\-sten\-ko, Dr.~A.Salpagarov and Dr.~O.Sumenkova
for the possibility of fruitful working under this problem. This
work has been partly supported by Russian Foundation of Basic
Research, grant 00--02--16108.

Authors thank an anonymous referee for useful remarks.

\end{acknowledgements}

\end{document}